\documentclass[a4paper,fleqn,usenatbib]{mnras}

\usepackage{newtxtext,newtxmath}
\usepackage[T1]{fontenc}
\usepackage{ae,aecompl}

\usepackage{graphicx}	% Including figure files
\usepackage{amsmath}	% Advanced maths commands
\usepackage{amssymb}	% Extra maths symbols

\title[Star formation-shock-AGN separation]{A New Diagnostic to Separate Line Emission from Star Formation, Shocks, and AGN Simultaneously in IFU Data}

\author[D'Agostino et al.]{
Joshua J. D'Agostino$^{1,2}$\thanks{E-mail: joshua.dagostino@anu.edu.au},
Lisa J. Kewley$^{1,2}$,
Brent A. Groves$^{1,2}$,
Anne Medling$^{3}$,
\newauthor
Michael A. Dopita,$^{1,2,4}$
Adam D. Thomas$^{1,2}$\\
$^{1}$Research School of Astronomy and Astrophysics, the Australian National University, Cotter Road, Weston, ACT 2611, Australia\\
$^{2}$ARC Centre of Excellence for All Sky Astrophysics in 3 Dimensions (ASTRO 3D)\\
$^{3}$Ritter Observatory, The University of Toledo, 2801 West Bancroft Street, Toledo, OH 43606 \\
$^{4}$Deceased
}

\date{Accepted XXX. Received YYY; in original form ZZZ}

\pubyear{2019}

\begin{document}
\label{firstpage}
\pagerange{\pageref{firstpage}--\pageref{lastpage}}
\maketitle

\begin{abstract}
In the optical spectra of galaxies, methods for the separation of line emission arising from star formation and an additional hard component, such as shocks or AGN, is well-understood and possible with current diagnostics. However, such diagnostics fail when attempting to separate and define line emission which arises from shocked gas, and that arising from AGN. We present a new three-dimensional diagnostic diagram for IFU data which can simultaneously separate the line emission amongst star formation, shocks, and AGN within a galaxy. We show that regions we define as AGN-dominated correlate well with the hard X-ray distribution in our test case NGC 1068, as well as with known regions of AGN activity in NGC 1068. Similarly, spaxels defined as shock-dominated correlate strongly with regions of high velocity dispersion within the galaxy. 
\end{abstract}


%%%%%%%%%%%%%%%%%%%%%%%%%%%%%%%%%%%%%%%%%%%%%%%%%%

%%%%%%%%%%%%%%%%% BODY OF PAPER %%%%%%%%%%%%%%%%%%

\section{Introduction} 
\label{sec:intro}

The mixing of line emission from gas ionised by star formation and active galactic nuclei (AGN) is a well-understood problem. The diagnostic diagram first described by \citet{BPT1981} (now known as the `BPT diagram') and subsequent diagnostic diagrams shown by \citet{VO1987} are powerful tools used in separating the emission from star formation and other excitation mechanisms. The BPT diagram in particular utilises the two emission line ratios [N \textsc{ii}]$\lambda 6584$/H$\alpha$ and [O \textsc{iii}]$\lambda 5007$/H$\beta$. The hard extreme ultraviolet (EUV) radiation field from the accretion disk of an AGN produces higher fluxes of collisionally-excited lines such as [N \textsc{ii}] and [O \textsc{iii}] \citep{Groves2004,Kewley2006,Kewley2013a}.

Initial work from \citet{Kewley2001} and subsequent work from \citet{Kewley2006}, \citet{Davies2014a,Davies2014b}, and \citet{TYPHOONpaper} has shown the existence of a `mixing sequence' amongst AGN galaxies, and within individual galaxies when using data from an integral field uniyt (IFU). Mainly using the BPT diagram, spaxels originating in the star-forming region of the BPT diagram demonstrate a continuous spread towards the AGN classification region, indicated by high [N \textsc{ii}]/H$\alpha$ and [O \textsc{iii}]/H$\beta$ ratios. This spread of spaxels on the BPT diagram has led to the name `starburst-AGN mixing' or `star formation-AGN mixing' to describe the process. IFU data is particularly helpful when studying star formation-AGN mixing within individual galaxies, due to the spatially-resolved spectrum it provides. 

The position of a galaxy spectra or spaxel on this mixing sequence can be used to determine the relative contribution of AGN. A problem with this method however, is that no other ionisation sources are considered. On the BPT diagram in particular, emission from shocked gas can produce [O \textsc{iii}]/H$\beta$ and [N \textsc{ii}]/H$\alpha$ ratios consistent with values along the mixing sequence, typically towards the AGN region of the diagram \citep[see][]{Rich2010,RKD2011,Kewley2013a}. This presents a problem with the results of star formation-AGN mixing, in that the ratio of true AGN emission within each spaxel is likely overestimated.

In this paper, we present a new diagnostic diagram capable of separating emission from star formation, shocks, and AGN simultaneously in IFU data. 

\section{Data selection}
\label{sec:data}

We use data from the Siding Spring Southern Seyfert Spectroscopic Snapshot Survey \citep[S7; ][]{S7}. The S7 is an IFU survey conducted between 2013 and 2016. The Wide Field Spectrograph \citep[WiFeS;][]{wifes1,wifes2} located on the ANU 2.3m telescope at Siding Spring Observatory was used for the survey. Data from the S7 is ideal to use for this work, due to the high spectral and spatial resolution of the survey. In particular, the high spectral resolution ($R \sim 7000$ in the red, 3000 in the blue) allows the study of independent velocity components in the emission lines, providing more information on shocked regions \citep{Ho2014}. For a full explanation of the S7, see \citet{S7}.

To showcase our new diagnostic diagram, we use the Seyfert 2 galaxy NGC 1068. NGC 1068 is the perfect galaxy to use as an example, as emission from all three sources of star formation, shocks, and AGN has been inferred within the nucleus. NGC 1068 has a large-scale biconical outflow structure, believed by many to be radiatively accelerated by the AGN \citep{Pogge1988,Cecil2002,Dopita2002b,TYPHOONpaper}. Observations have also shown evidence of shocked gas towards the edge of the central bar \citep{Tacconi1994}. Evidence of shocks within the system is further supported by the velocity of the outflowing material, calculated to be moving away from the nucleus at velocities in excess of 3000 kms${-1}$ \citep{Cecil2002}. Furthermore, NGC 1068 has been shown to host a circumnuclear star-forming ring, with estimates of a star formation rate (SFR) of up to 100 $M_\odot$ yr$^{-1}$ \citep{Thronson1989}. 

\section{Issues with current diagnostics}
\label{sec:issues}

\begin{figure}
\centering
\includegraphics[width=\columnwidth]{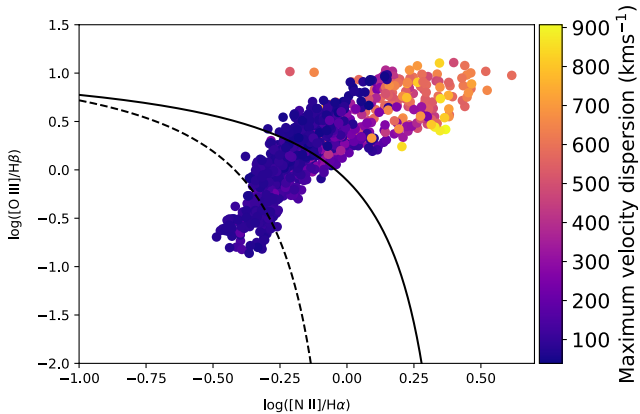}
\caption{BPT diagram of NGC 1068, coloured by the maximum velocity dispersion in each spaxel.}
\label{fig:1068bpt}
\end{figure}
 
In Figure~\ref{fig:1068bpt} we show a BPT diagram of NGC 1068, coloured by the maximum velocity dispersion in the spaxel. The maximum velocity dispersion in each spaxel is calculated by firstly fitting each emission line with three Gaussian components \citep[see][]{LZIFU}. The recommended number of one, two, or three Gaussian components for each spaxel is then determined by a neural network \citep[\textsc{LZComp};][]{Hampton2017}. The maximum velocity dispersion in each spaxel is then defined as the velocity dispersion of the highest-order component. We refer to the velocity dispersion of each individual component as the `single-component velocity dispersion' later in Section~\ref{sec:3ddiag}.

We show the maximum velocity dispersion in each spaxel in Figure~\ref{fig:1068bpt} to highlight the prevalence of shocked gas along and within the mixing sequence of the galaxy. The spaxels on the diagram form a spread from low to high emission line ratios, indicating a large contribution from the AGN. However, many of the spaxels found above the \citet{Kewley2001} line show velocity dispersions greater than 300 kms$^{-1}$. At these velocity dispersions and higher, fast shocks very likely contribute to the ionisation present in these spaxels \citep[e.g.][]{Dopita1995,RKD2011,Ho2014}. Yet the emission line ratios of the spaxels are high enough to be consistent with Seyfert emission \citep{Kewley2006}. Hence, Figure~\ref{fig:1068bpt} shows that the BPT diagram is poor at successfully separating emission between AGN and shocks. 

\section{The 3D diagnostic diagram}
\label{sec:3ddiag}

\begin{figure*}
\centering
\includegraphics[width=\textwidth]{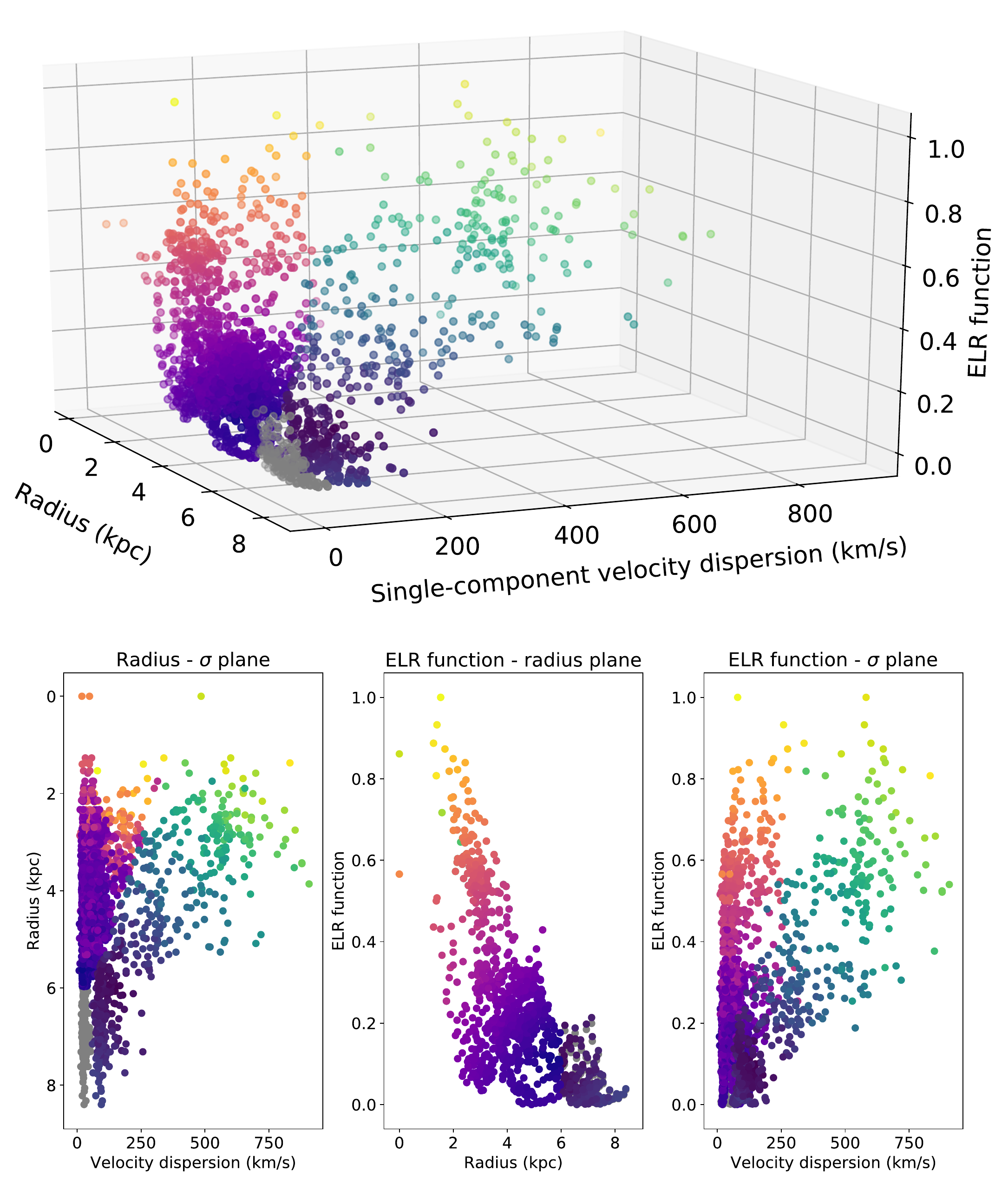}
\caption{3D diagnostic diagram of NGC 1068, showing two distinct mixing sequences of spaxels. The velocity dispersion value is the velocity dispersion of the individual components. Each individual-component velocity dispersion is combined with the total flux (`zeroth' component), and the radius value for the spaxel to form a data point. The purple-to-yellow sequence is referred to as the `first' sequence, and the deep blue-to-yellow sequence is referred to as the `second' sequence. Grey spaxels are those which are not definitively in either sequence. The first sequence shows mixing between emission from star formation and AGN, and the second sequence shows the mixing between the star formation and shocks. Significant scatter appears between the two sequences, indicating mixing also between the AGN and shocked emission. The three panels below the 3D diagram show the projection in the radius-velocity dispersion plane, radius-ELR function plane, and velocity dispersion-ELR function plane. The separation into two sequences is seen more clearly when studying all three planes.}
\label{fig:3ddiag}
\end{figure*}

\begin{figure}
\centering
\includegraphics[width=\columnwidth]{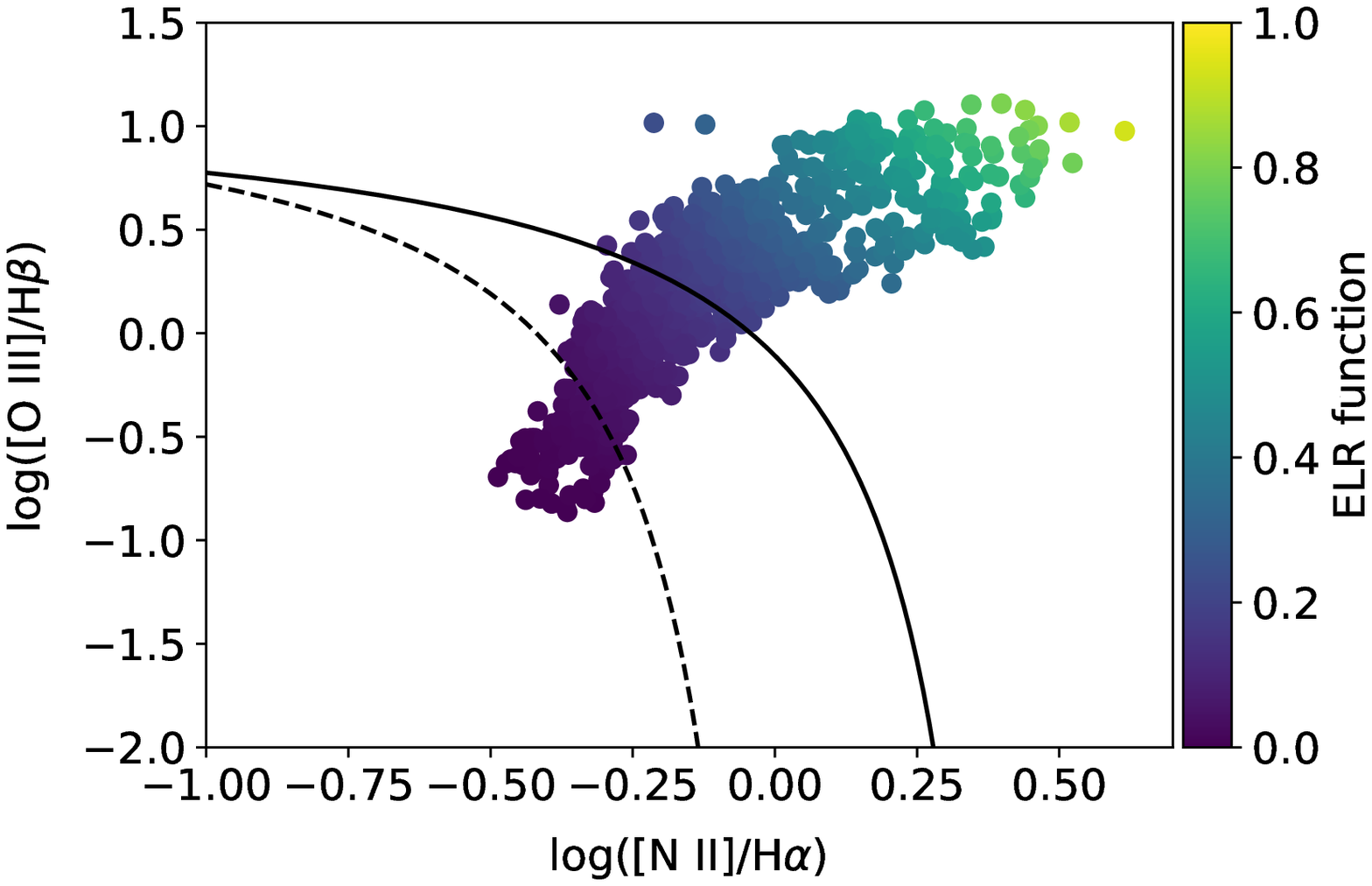}
\caption{BPT diagram of NGC 1068, coloured by the ELR function value from Equation~\ref{eq:elr} in each spaxel. The black, solid line is the \citet{Kewley2001} demarcation line, and the black, dashed line is the \citet{Kauffmann2003} demarcation line.}
\label{fig:elrfunc}
\end{figure}

We present a new three-dimensional diagnostic diagram to simultaneously separate emission amongst star formation, shocks, and AGN in IFU data. This diagram is shown for NGC 1068 in Figure~\ref{fig:3ddiag}. The emission line ratio information from the BPT diagram is still considered in this diagram, shown on the vertical axis as our emission line ratio (ELR) function. The functional form is given in Equation~\ref{eq:elr}. The ELR function essentially traces the mixing sequence of the galaxy, shown in Figure~\ref{fig:elrfunc}. The purpose of the ELR function is to order the spaxels on the 3D diagram in terms of their combined [O \textsc{iii}]/H$\beta$ and [N \textsc{ii}]/H$\alpha$ ratios. It is important to note that the ELR function is data-dependent, and the range of values from 0 to 1 is arbitrary. The endpoints of 0 and 1 do not represent any physical phenomena, such as 100\% star formation emission and 100\% AGN emission respectively. The other axes of radius and velocity dispersion provide additional information on the data. As mentioned previously, the velocity dispersion is a suitable shock diagnostic. Furthermore, the radial values provide information on the mixing between AGN and other processes, as AGN emission is seen more towards the nucleus. In addition to this, when mixing with other processes, star formation is more likely seen towards the outskirts of the galaxy \citep[e.g.][]{Davies2014a,Davies2014b,TYPHOONpaper}. The single-component velocity dispersion used in Figure~\ref{fig:3ddiag} refers to the velocity dispersion of individual components in each spaxel. If the neural network \textsc{LZComp} recommends multiple-component fits to a spaxel, then multiple data points for a single spaxel will be present on the 3D diagram. These data points will differ in their values for the velocity dispersion (with higher-order components having a greater velocity dispersion), yet will have the same radial and ELR function values. The ELR function value associated with these multiple components of a single spaxel is calculated from the total flux (`zeroth' component) of the spaxel. We use the single-component velocity dispersion on the 3D diagram, because the various components contain information about different processes. The first component (which contains the lowest velocity dispersions) is consistent with emission from H \textsc{ii} regions, and the higher-order components (in particular, the third component if present) is consistent with emission from shocks \citep{Ho2014}.

\begin{equation}
\begin{aligned}
\mathrm{ELR \;function} = \frac{\mathrm{log([N \textsc{ii}]/H}\alpha) - \mathrm{min}_{\mathrm{log([N \textsc{ii}]/H}\alpha)}}{\mathrm{max}_{\mathrm{log([N \textsc{ii}]/H}\alpha)} - \mathrm{min}_{\mathrm{log([N \textsc{ii}]/H}\alpha)}}\\
 \times \frac{\mathrm{log([O \textsc{iii}]/H}\beta) - \mathrm{min}_{\mathrm{log([O \textsc{iii}]/H}\beta)}}{\mathrm{max}_{\mathrm{log([O \textsc{iii}]/H}\beta)} - \mathrm{min}_{\mathrm{log([O \textsc{iii}]/H}\beta)}}
\end{aligned}
\label{eq:elr}
\end{equation}

\section{Results}
\label{sec:res}

The 3D diagnostic diagram demonstrates two distinct sequences of spaxels, each advancing from a region of low ELR function values, to regions of high ELR function values. The two sequences separated by colour on Figure~\ref{fig:3ddiag}. Hereafter, we refer to the purple-to-yellow sequence as the `first' sequence, and the deep blue-to-yellow sequence as the `second' sequence. Grey spaxels are those which do not clearly belong to either sequence. The separation of the spaxels into two sequences is seen more clearly in the bottom panels of Figure~\ref{fig:3ddiag}, which shows the two sequences as viewed in the radius-velocity dispersion plane, radius-ELR function plane, and velocity dispersion-ELR function plane. In general, we find that the first and second sequences mostly contain spaxels of the first-component and third-component velocity dispersions respectively. However, a small fraction of first-component velocity dispersion spaxels are located in the second sequence. Second-component velocity dispersion spaxels are found within and between both sequences. Large scatter is seen between the two sequences, particularly at high ELR function values. Therefore, it should be noted that the separation made between the two sequences is only approximate.

\begin{figure*}
\centering
\includegraphics[width=\textwidth]{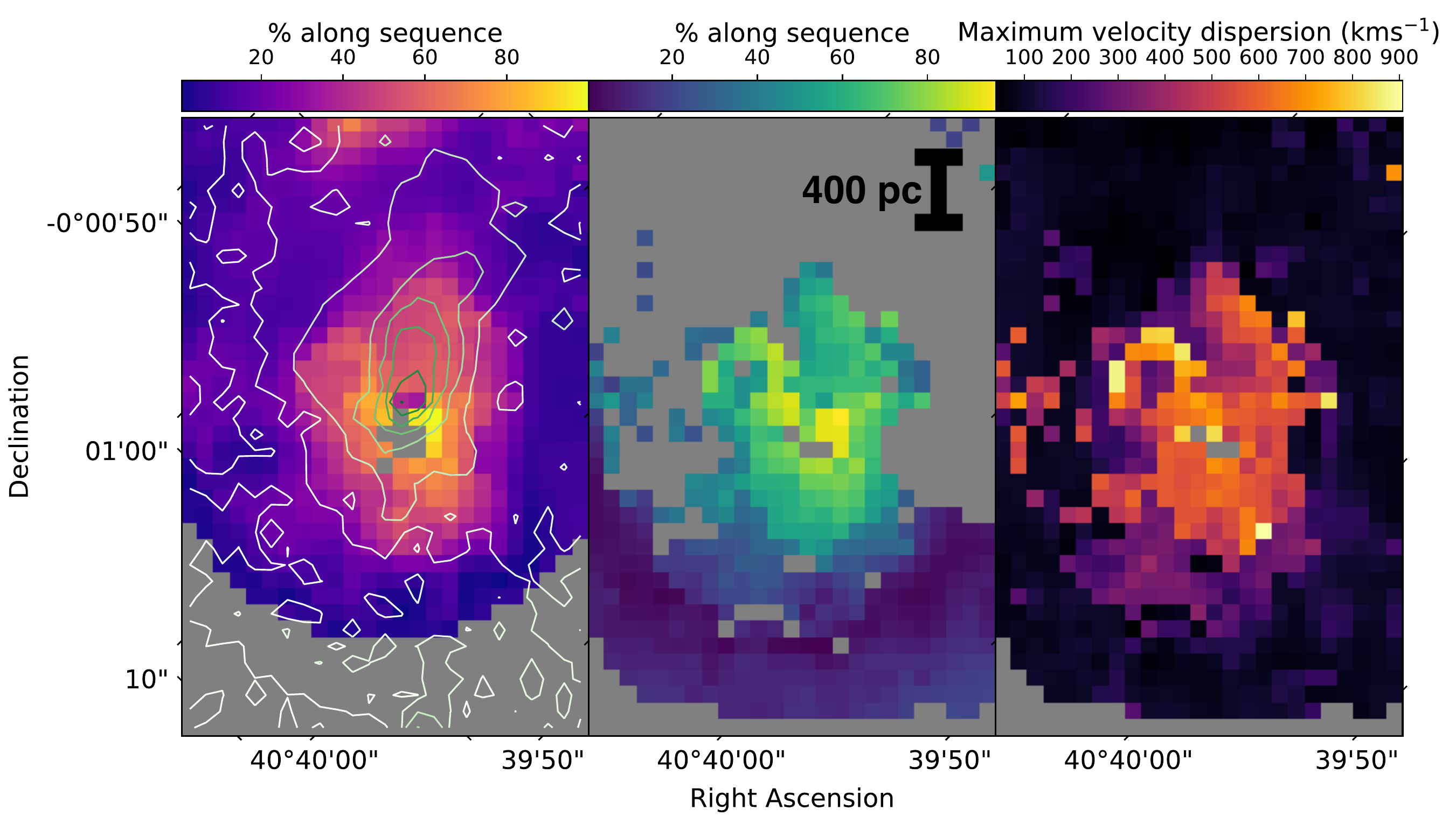}
\caption{Maps of NGC 1068, showing the distribution of the first sequence of spaxels in (a), the second sequence in (b), and the velocity dispersion in (c). Spaxel colours for (a) and (b) are identical to Figure~\ref{fig:3ddiag}. 0.25-7.5 keV \textit{Chandra} X-ray contours from \citet{Young2001} are also shown on (a).}
\label{fig:3maps}
\end{figure*}

We claim that the first and second sequences represent the star formation-AGN mixing and star formation-shock mixing respectively. We show this by studying the spatial distribution of the two sequences in Figure~\ref{fig:3maps}. The map of the first sequence in Figure~\ref{fig:3maps}a closely traces the 0.25-7.5 keV \textit{Chandra} X-ray contours from \citet{Young2001}. Furthermore, the distribution of the first-sequence spaxels in Figure~\ref{fig:3maps}a resembles the [O \textsc{iii}] flare in NGC 1068, believed to be a result of photoionisation from the AGN. The alignment with the X-ray and [O \textsc{iii}] distributions is a strong indication that the first sequence represents the star formation-AGN mixing in each spaxel. Figures~\ref{fig:3maps}b and~\ref{fig:3maps}c however show a strong correlation between the second sequence of spaxels and the velocity dispersion distribution. Spaxels found towards the high ELR end of the second sequence also contain extremely high velocity dispersions. The combination of high ELR and velocity dispersion values indicate the emission in these spaxels is from shocked gas. Significant scatter between the two sequences exists at high ELR values, seen in Figure~\ref{fig:3ddiag}, indicating mixing also between the AGN- and shock-affected spaxels. This is to be expected however, as shocks can result from the accretion disk of the AGN \citep[e.g.][]{Spruit1987,Molteni1994,SM1994}.

\begin{figure}
\centering
\includegraphics[width=\columnwidth]{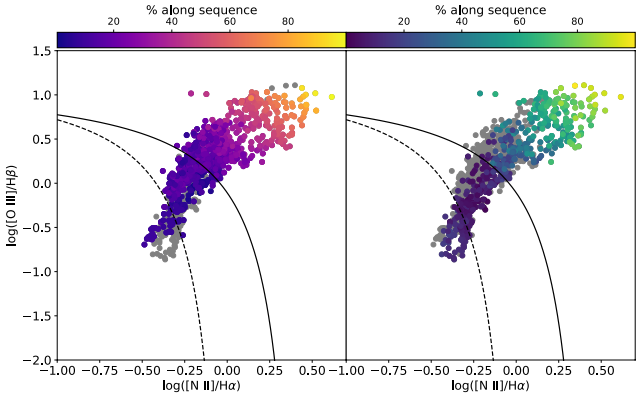}
\caption{BPT diagrams showing the spread of spaxels in the first sequence on the left, and second sequence on the right. Spaxel colours are identical to Figure~\ref{fig:3ddiag}. The black, solid line is the \citet{Kewley2001} demarcation line, and the black, dashed line is the \citet{Kauffmann2003} demarcation line.}
\label{fig:seqsbpt}
\end{figure}

The star formation-AGN and star formation-shock mixing nature of the two sequences is also shown in Figure~\ref{fig:seqsbpt}. Both sequences begin in the pure star-forming region of the BPT diagram \citep[below the demarcation line of ][]{Kauffmann2003} before demonstrating a continuous spread towards and beyond the \citet{Kewley2001} demarcation line. The BPT diagrams show that both sequences contain mixing between pure star formation and additional hard components, demonstrated prior to be predominantly AGN and shocks for the first and second sequences respectively.
% bpts

\section{Summary and future work}
\label{sec:summary}

We have shown that traditional methods used to distinguish non-stellar line emission are ineffective in separating AGN- and shock-excited gas. Instead, we have proposed and demonstrated a new three-dimensional diagnostic diagram which simultaneously separates emission from star formation, shocks, and AGN in IFU data. The diagram uses the radial information and velocity dispersion in each spaxel for two axes, while the third axis consists of a function of emission line ratios. The emission line ratio function used in the diagram is a combination of the BPT emission line ratios [N \textsc{ii}]$\lambda 6584$/H$\alpha$ and [O \textsc{iii}]$\lambda 5007$/H$\beta$, ensuring that the advantages of the BPT diagram are still retained.

Using the Seyfert galaxy NGC 1068 as a test case, we show that two distinct and clear sequences of spaxels are found on the 3D diagram. We claim these sequences represent the mixing between star formation and AGN, and star formation and shocks respectively. This is supported through maps of the spaxel distributions across the galaxy. The distribution of spaxels found in the star formation-AGN mixing sequence correlates well with the X-ray distribution in NGC 1068, as well as the [O \textsc{iii}] flare seen extending from the galaxy. The spaxels within the star formation-shock mixing sequence however correlate strongly with regions of high velocity dispersion in the galaxy. These results suggest the two sequences seen on the 3D diagnostic diagram indeed represent the star formation-AGN and star formation-shock mixing within the galaxy.

In an upcoming publication, we intend to use this method to quantify the relative contributions of star formation, shocks, and AGN to emission lines for galaxies undergoing mergers, as a function of merger stage. 

%In an upcoming publication, we intend to use the 3D diagnostic diagram to calculate and quantify the relative contributions of star formation, shocks, and AGN to the emission line ratios for .

\section*{Acknowledgements}

Parts of this research were conducted by the Australian Research Council Centre of Excellence for All Sky Astrophysics in 3 Dimensions (ASTRO 3D), through project number CE170100013. 

%%%%%%%%%%%%%%%%%%%%%%%%%%%%%%%%%%%%%%%%%%%%%%%%%%



%\bibliographystyle{mnras}

\bibliographystyle{mnras}
\bibliography{3dletter}

%\begin{thebibliography}{99}
%\end{thebibliography}

%%%%%%%%%%%%%%%%%%%%%%%%%%%%%%%%%%%%%%%%%%%%%%%%%%

%%%%%%%%%%%%%%%%% APPENDICES %%%%%%%%%%%%%%%%%%%%%

%%%%%%%%%%%%%%%%%%%%%%%%%%%%%%%%%%%%%%%%%%%%%%%%%%

% Don't change these lines
\bsp	% typesetting comment
\label{lastpage}
\end{document}